\documentclass[floatfix,pre,showpacs]{revtex4}
\usepackage{graphicx}
\usepackage{amsmath}
\usepackage{latexsym}
\begin{document}

\title{Nematic cells with defect--patterned alignment layers}

\author{Adam~S. Backer}
\affiliation{Department of Physics, Brown University, Providence,
Rhode Island 02912}
\author{A.~C. Callan--Jones}
\altaffiliation[Present address:  ]
{Institut Curie, UMR CNRS 168, 26 rue d'Ulm 75248 Paris Cedex 05, France}
\author{Robert A. Pelcovits}
\affiliation{Department of Physics, Brown University, Providence,
Rhode Island 02912}

\date{\today}
\begin{abstract}
Using Monte Carlo simulations of the Lebwohl--Lasher model we study the director ordering in a nematic cell where the top and bottom surfaces are patterned with a lattice of $\pm 1$ point topological defects of lattice spacing $a$. We find that the nematic order depends crucially on the ratio of the height of the cell $H$ to $a$. When $H/a \gtrsim 0.9$ the system is very well--ordered and the frustration induced by the lattice of defects is relieved by a network of half--integer defect lines which emerge from the point defects and hug the top and bottom surfaces of the cell. When $H/a \lesssim 0.9$  the system is disordered and the half--integer defect lines thread through the cell joining point defects on the top and bottom surfaces. We present a simple physical argument in terms of the length of the defect lines to explain these results. To facilitate eventual comparison with experimental systems we also simulate optical textures and study the switching behavior in the presence of an electric field.
\end{abstract}
\maketitle
\section{Introduction}
\begin{figure}[h]
\begin{center}
\includegraphics[width=5.0in,clip=true]{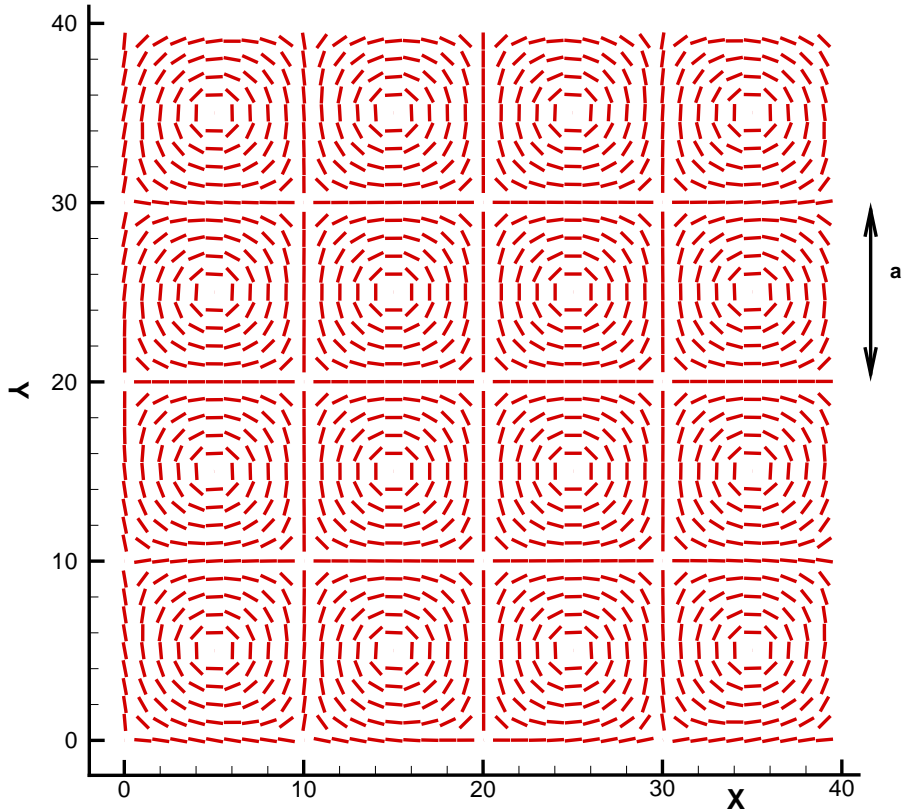}
\caption{(Color online) Schematic of a patterned alignment layer created with linear photopolymerizable polymers and a suitable holographic interference pattern. Nematic molecules brought into contact with this surface will align with the rods which indicate the local direction of the polymer chains. This pattern consists of two interpenetrating square lattices of +1 and -1 point defects with lattice spacing $a=10$. The +1 and -1 defects are located at coordinates: $x=n\frac{a}{2},y=m\frac{a}{2};\ \ n,m=1,3,5...$, and $x=ka,y= la;\ \ k,l=0,1,2,...$ respectively. }
\label{pattern}
\end{center}
\end{figure}

The use of nematic liquid crystals in display devices requires alignment of the liquid crystal on the surfaces that bound the top and bottom of the display cell. Traditionally this alignment is achieved by rubbing the polyimide alignment layers with a velvet cloth. However, this approach leads to contamination and thus there has been considerable effort devoted to achieving alignment via noncontact means, among them the use of linear photopolymerizable polymers \cite{Schadt:96,Chen:95,Wen:02}. One way to achieve director alignment with the polymer is to irradiate it with UV light which polymerizes the material along the local direction of polarization of the light \cite{Gibbons:91,Lee:05,Crawford:05}. When the nematic material comes in direct contact with the polymer layer subsequent to the UV exposure, the nematic director is planar aligned locally along the direction of the polymer chains. Not only does this alignment process reduce contamination, it also allows for the creation of spatially modulated alignment patterns when the polymeric material is exposed to holographically generated polarization interference patterns. Depending on the optics used to generate the interference pattern the modulation can be either one \cite{Eakin:04} or two dimensional \cite{Gorkhali:thesis} in nature. The two-dimensional patterns include an ordered array of point topological defects shown in Fig.~\ref{pattern}, where +1 and -1 defects occupy interpenetrating square lattices of lattice constant \textit {a} \cite{Crawford:private}.  While the net topological charge of the alignment layer shown in the figure is zero, the pattern introduces frustration and one can ask what the nature of the nematic order in a cell whose alignment surfaces on the top and bottom look like this. Energy considerations require that the point defects on the alignment layers lie on defect lines that penetrate into the nematic. Do these lines thread through the cell joining point defects on opposite sides, or is another geometry possible? Can this cell exhibit net nematic order given the frustration induced by the boundaries? In this paper we answer these questions using Monte Carlo simulations. We find that the ratio of the height of the cell (i.e., the distance between the patterned alignment layers, often referred to as the ``cell gap'') and the lattice spacing $a$ between the topological defects plays a crucial role in determining the degree of nematic order and the geometry of the defect lines. 

\section{Simulations}
\label{simulation}
We model the nematic cell using the Lebwohl--Lasher model \cite{Lasher:72}, a lattice model of rotors
 specified by unit vectors ${\mathbf s}_i$ located at sites $i$ of a cubic lattice. Each rotor represents a small group of mesogenic molecules. The model is very well suited to numerical studies of the director structure of nematic liquid crystals. It has been intensively investigated using Monte Carlo
techniques since its introduction
\cite{Luckhurst:82,Fabbri:86,Zhang:92,Allen:05}, including applications to nematic cells and display devices \cite{Chiccoli:98,Chiccoli:01}.
The model is defined by the Hamiltonian:

\begin{equation}
{\cal {H}}= -J \sum _{<ij>}\biggl\lbrace {3 \over 2}
({\mathbf s}_{i} \cdot  {\mathbf s}_{j})^2 -{1\over
2}\biggr\rbrace
\end{equation}
where the sum is over nearest neighbors and $J$ is a coupling
parameter. 
 We consider a nematic cell of size $L\times L$ in the $x-y$ plane and height $H$ along the $z$ axis. The patterned alignment layers are located at $z=0$ and $z=H-1$, and we assume infinitely strong anchoring of the rotors on these layers. The two alignment layers, each of the structure shown in Fig.~\ref{pattern}, are assumed to be identical and in registry with each other, consistent with the experimental fabrication process where the cell is assembled prior to the holographic exposure and introduction of the nematic material \cite{Crawford:05}. Periodic boundary conditions are imposed on the remaining four faces of the cell.  We carried out our simulations at a temperature $T=0.1$, measured in dimensionless units of $ J/k_B $ . We initialized the system with a random configuration of rotor orientations (consistent with the experimental procedure where the nematic is introduced into the cell in the isotropic phase and then cooled) and ran simulations for $50 000$ Monte Carlo cycles where each cycle corresponds to attempted rotations of $N=L^2 H$ rotors chosen at random. Our largest systems equilibrated in less than $25 000$ cycles. We implemented the attempted rotations via random displacements of $\phi$ and $\cos\theta$, where $\phi$ and $\theta$ are the spherical coordinates of a rotor. The ranges of these random displacements were chosen to be: $-0.025<\cos\theta<0.025$ and $-0.25<\phi<0.25$. These choices allowed us to achieve a Monte Carlo acceptance ratio of approximately $50\%$.

\begin{figure}
\includegraphics[height=3in]{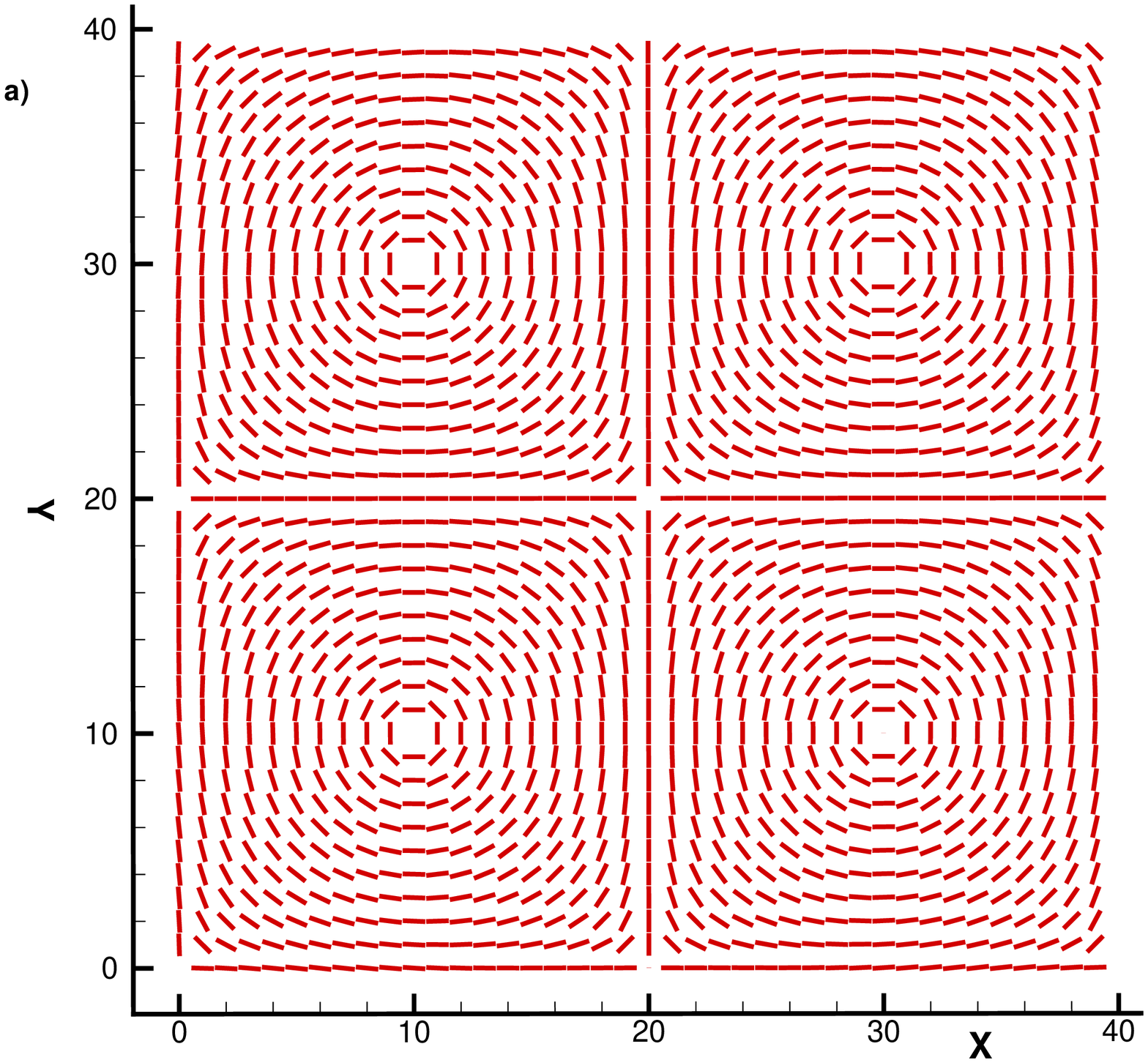}
\includegraphics[height=3in]{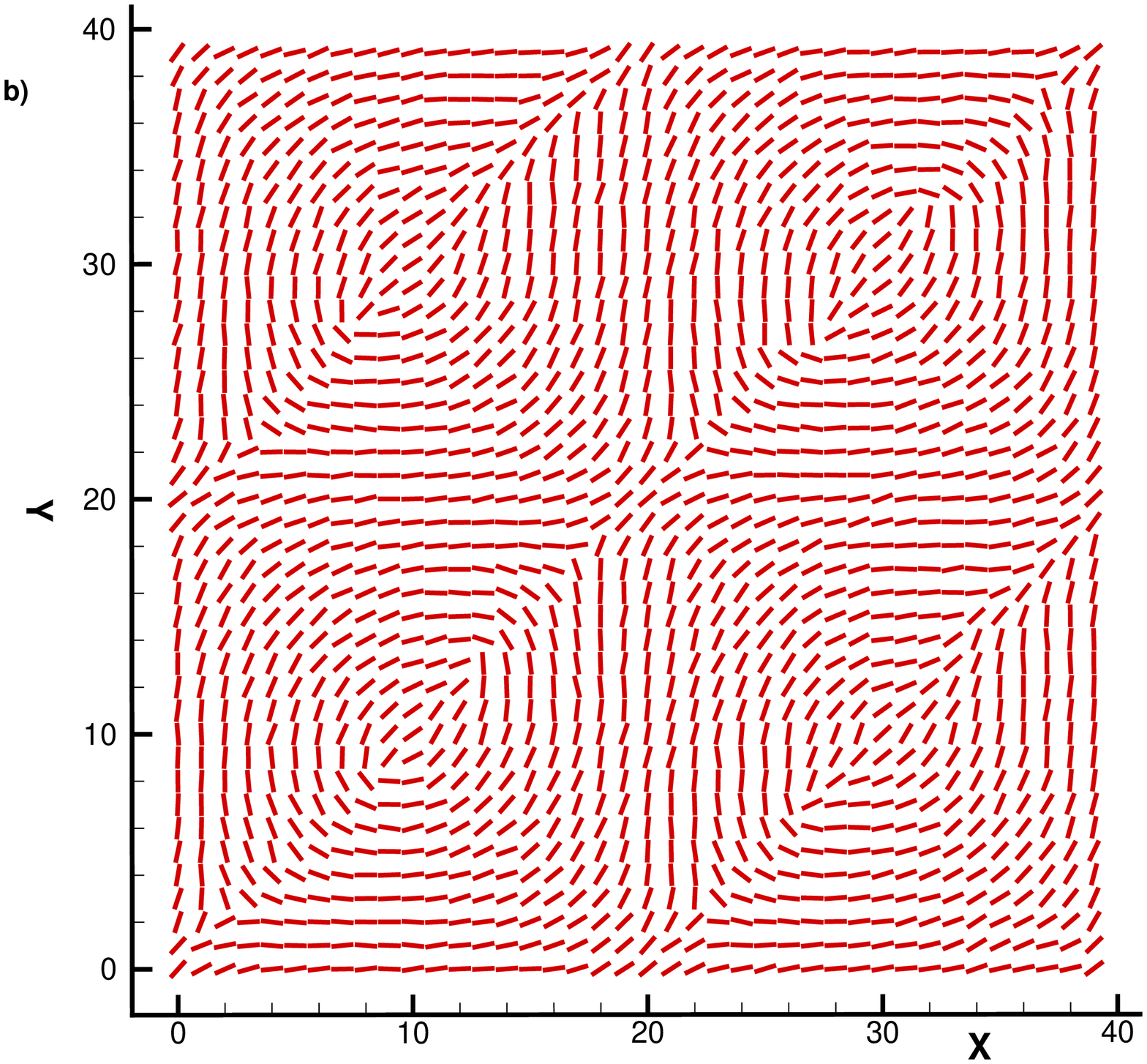}
\includegraphics[height=3in]{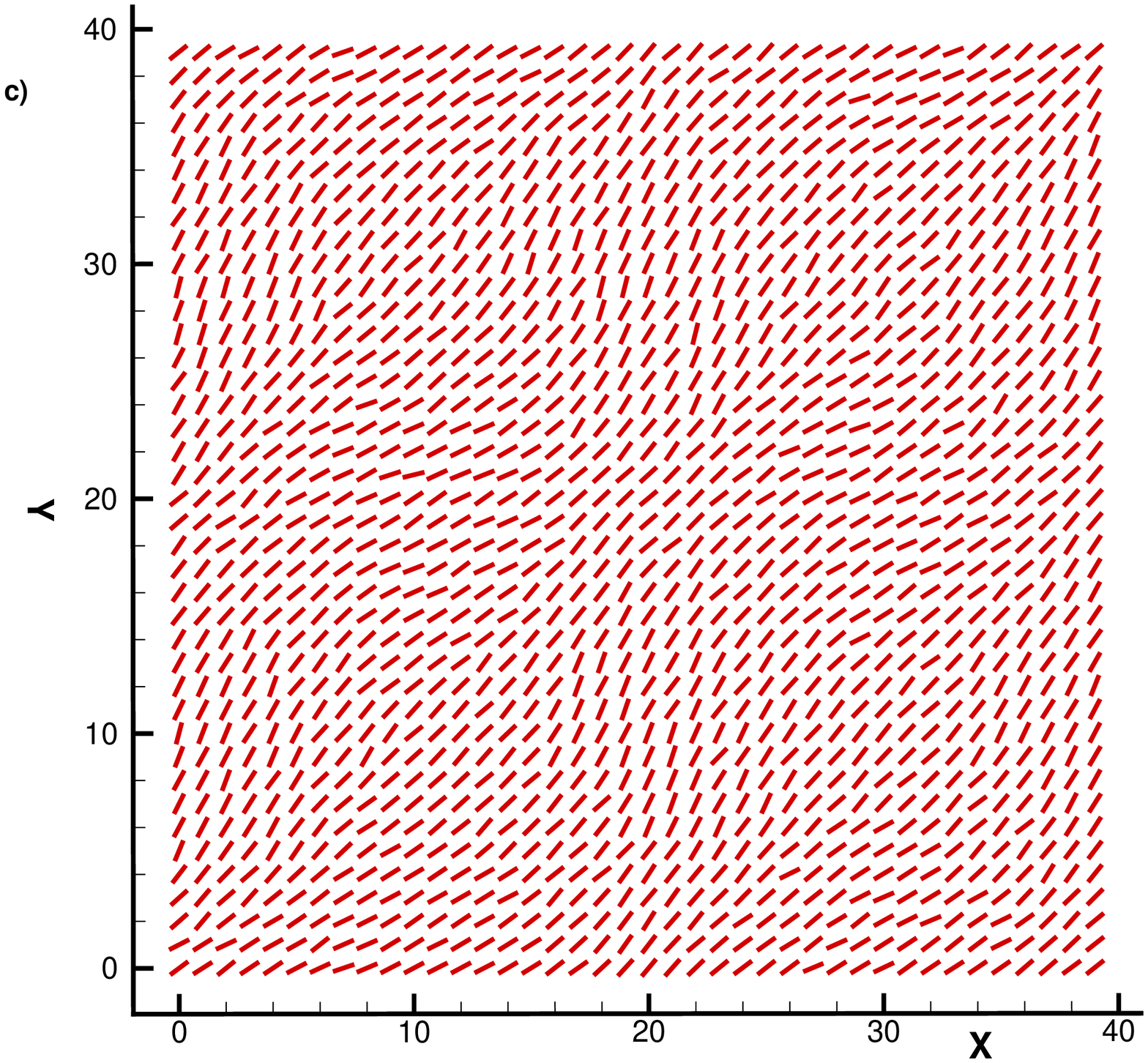}
\includegraphics[height=3in]{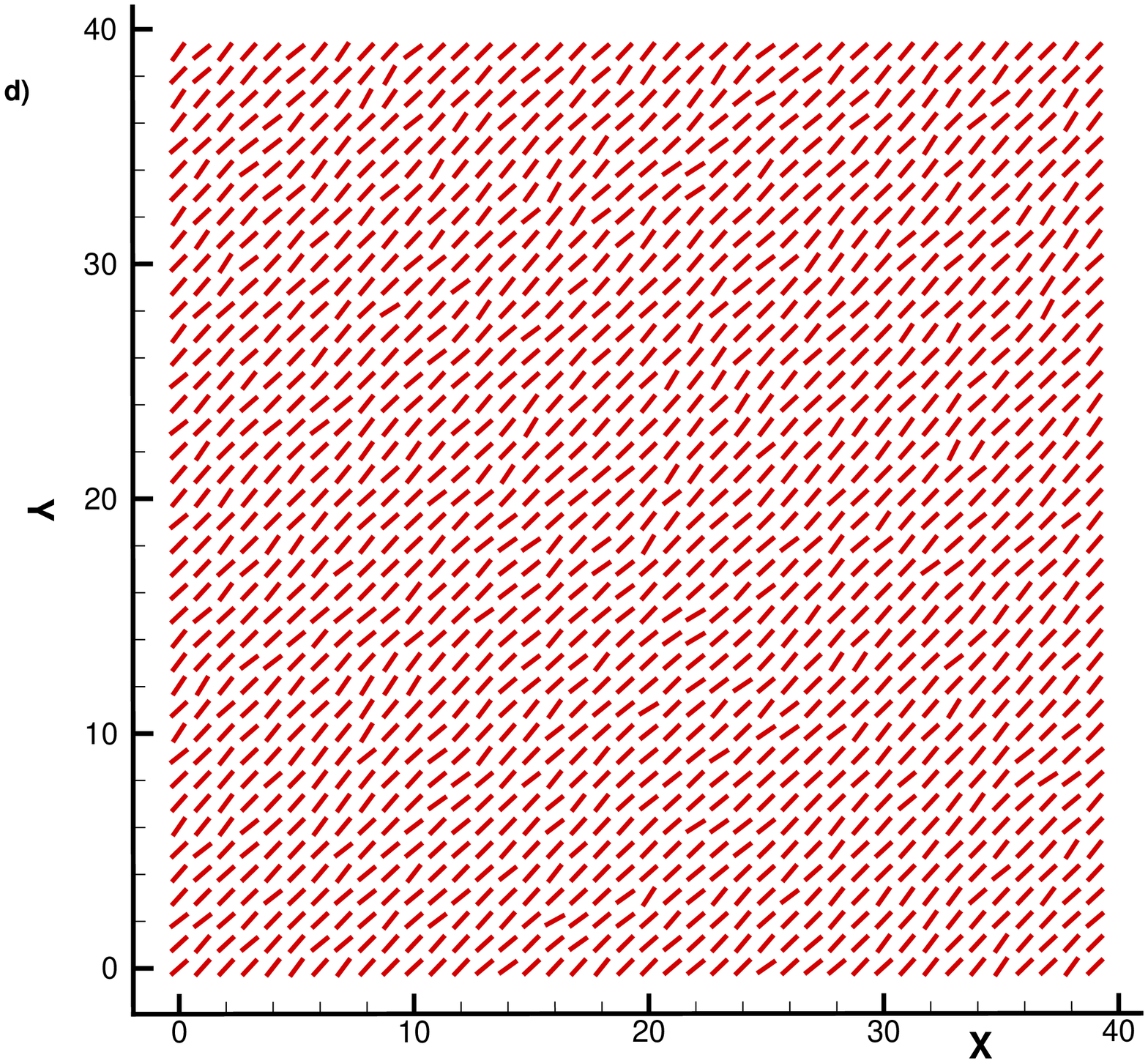}
\caption{(Color online) Director patterns in three layers of a cell of size $L=40$, $H=40$ with defect lattice spacing $a=20$. The two identical patterned layers ($z=0,39$) are displayed in (a). Layer $z=1$, shown in (b), illustrates the bifurcation of the integer--valued point defects on the patterned surfaces into pairs of half--integer defect lines. Deeper in the cell ((c), $z=4$, (d), $z=20$), the defect lines are not present and very uniform nematic order exists along a diagonal direction in the $x-y$ plane.}
\label{largeH}
\end{figure}

\begin{figure}
\includegraphics[height=3in]{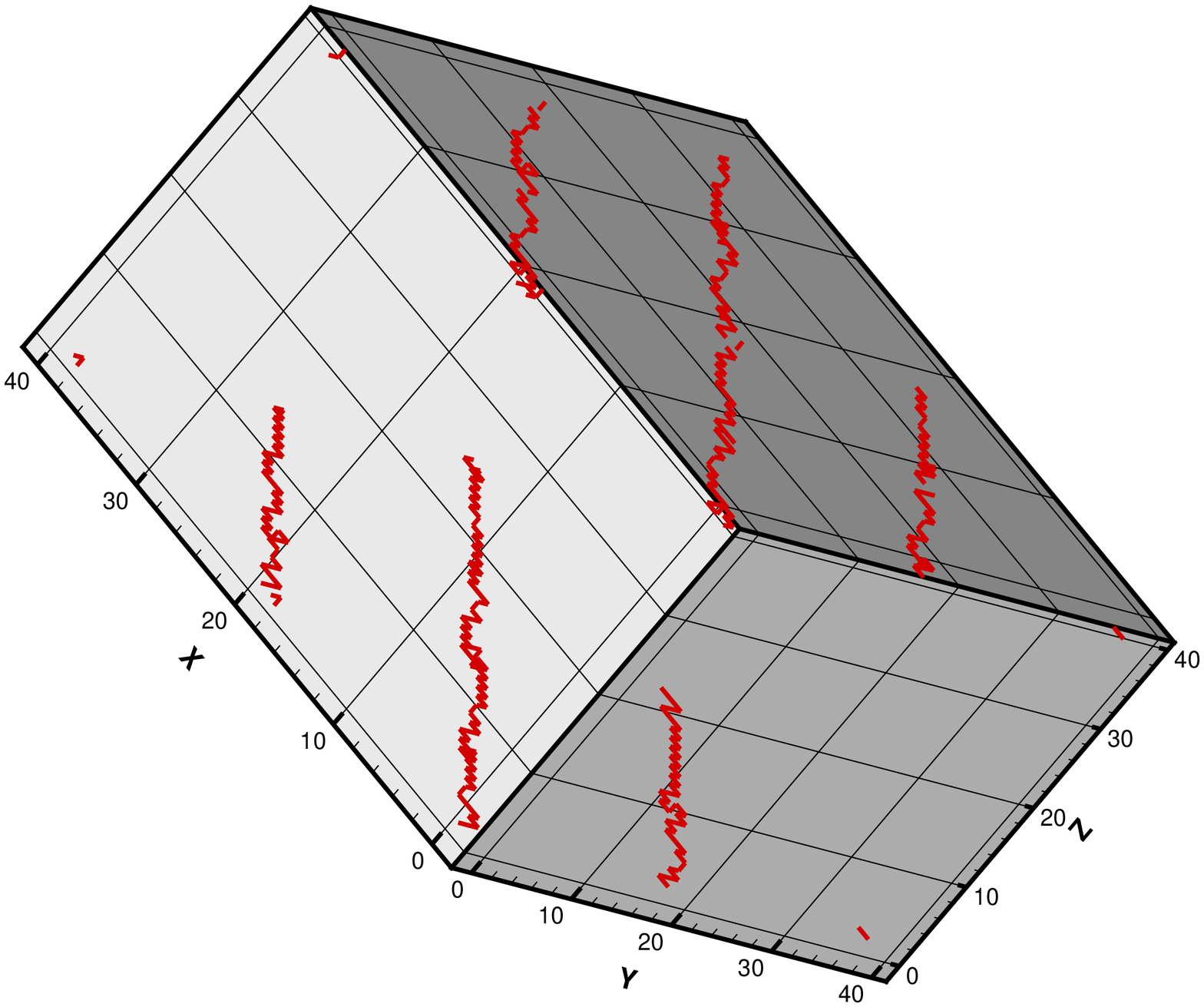}
\includegraphics[height=3in]{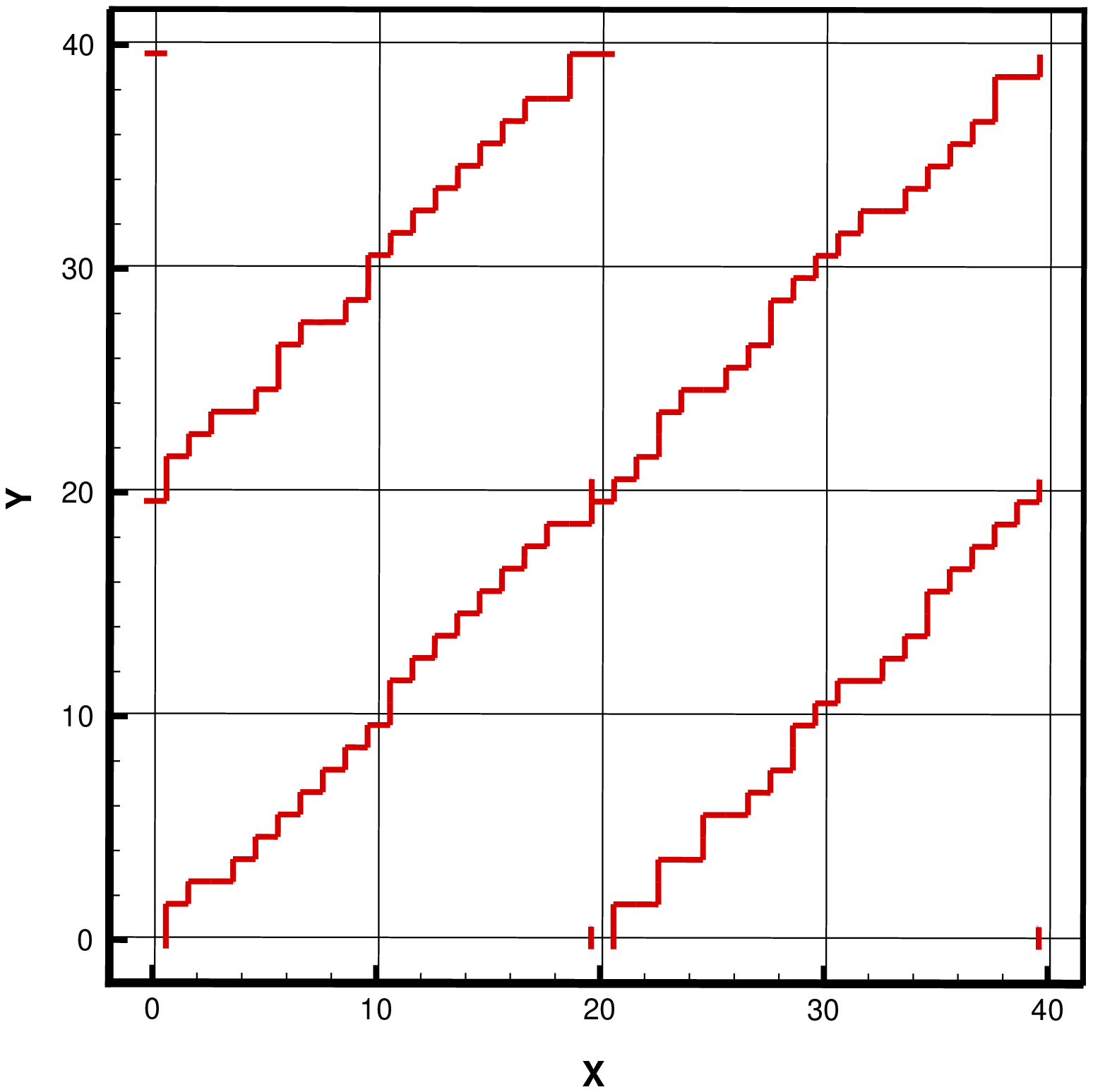}
\caption{(Color online) Half--integer valued defect lines in a cell of size $L=40$, $H=40$ with defect lattice spacing $a=20$. The lines have been identified using the method of Ref.~\cite{Zapotocky:95}. The full cell is shown in (a) and a topview (looking down the $z$ axis) is shown in (b). The defect lines emerge from the integer--valued point defects on the patterned boundary layers located at $z=0,39$. Each defect line joins a pair of oppositely charged point defects on the same boundary layer. The nematic is ordered along the same diagonal direction followed by the defect lines shown in (b); compare with Fig.~\ref{largeH}d.}
\label{largeHdefects}
\end{figure}

\begin{figure}
\includegraphics[height=3in]{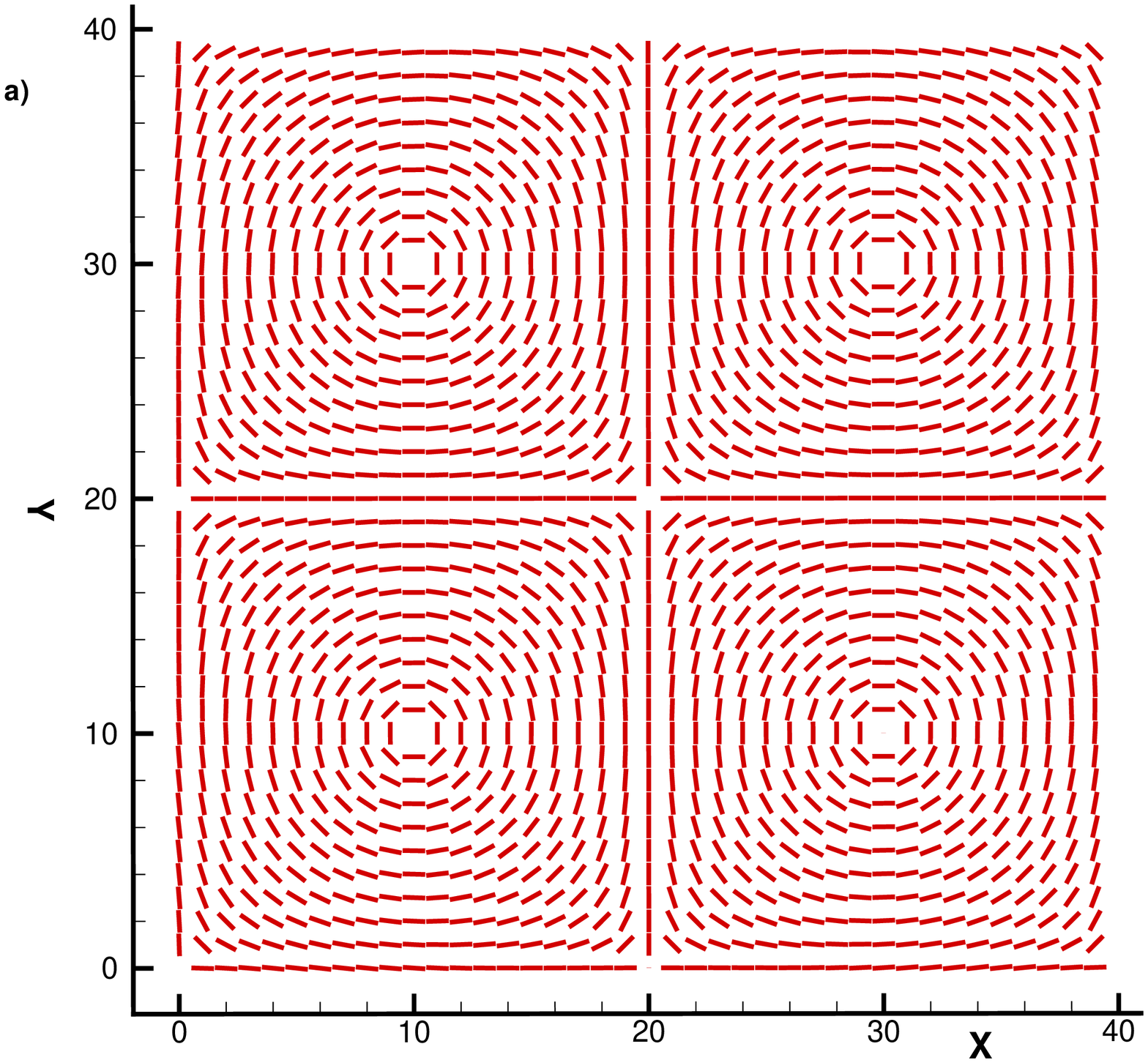}
\includegraphics[height=3in]{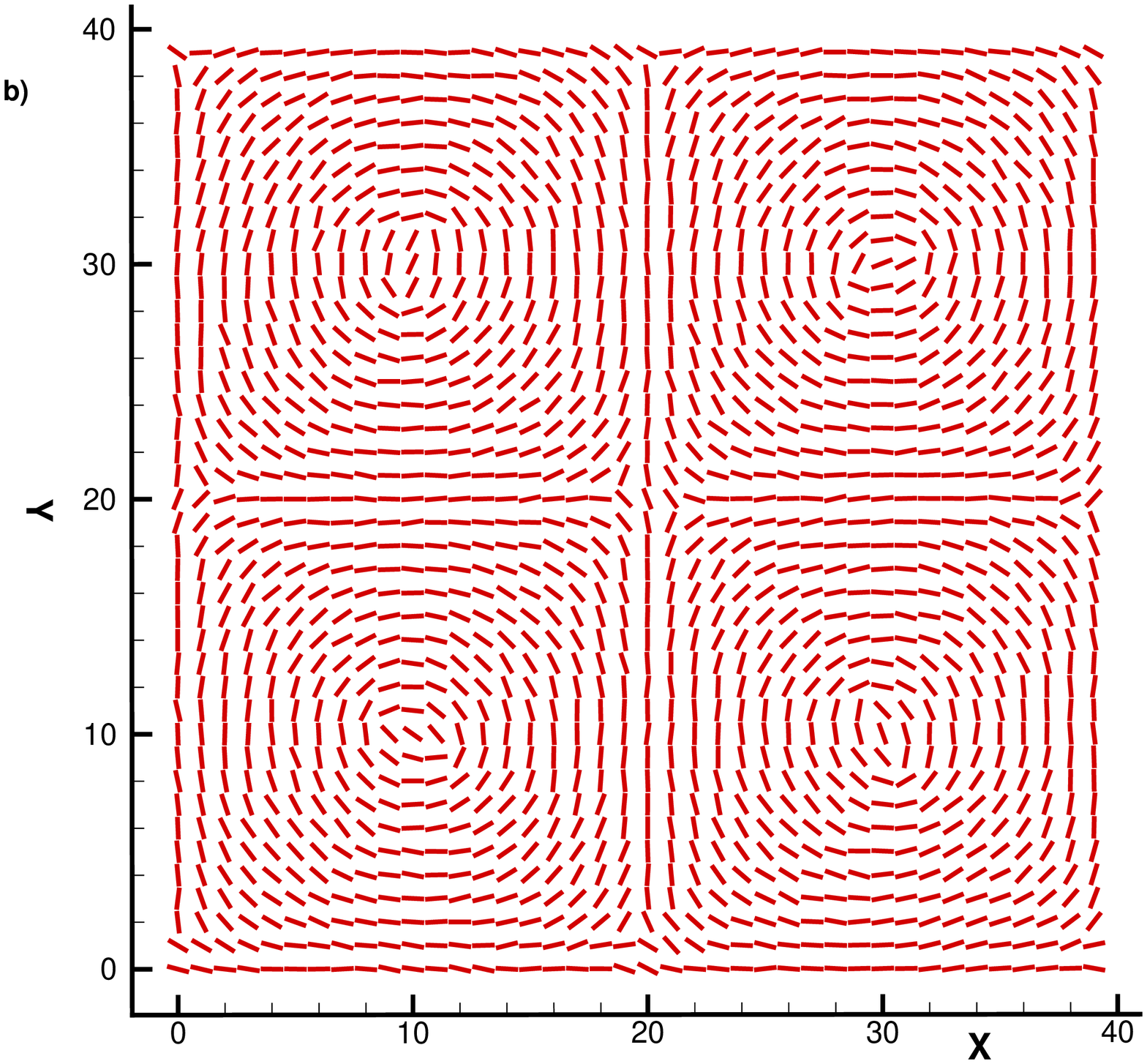}
\includegraphics[height=3in]{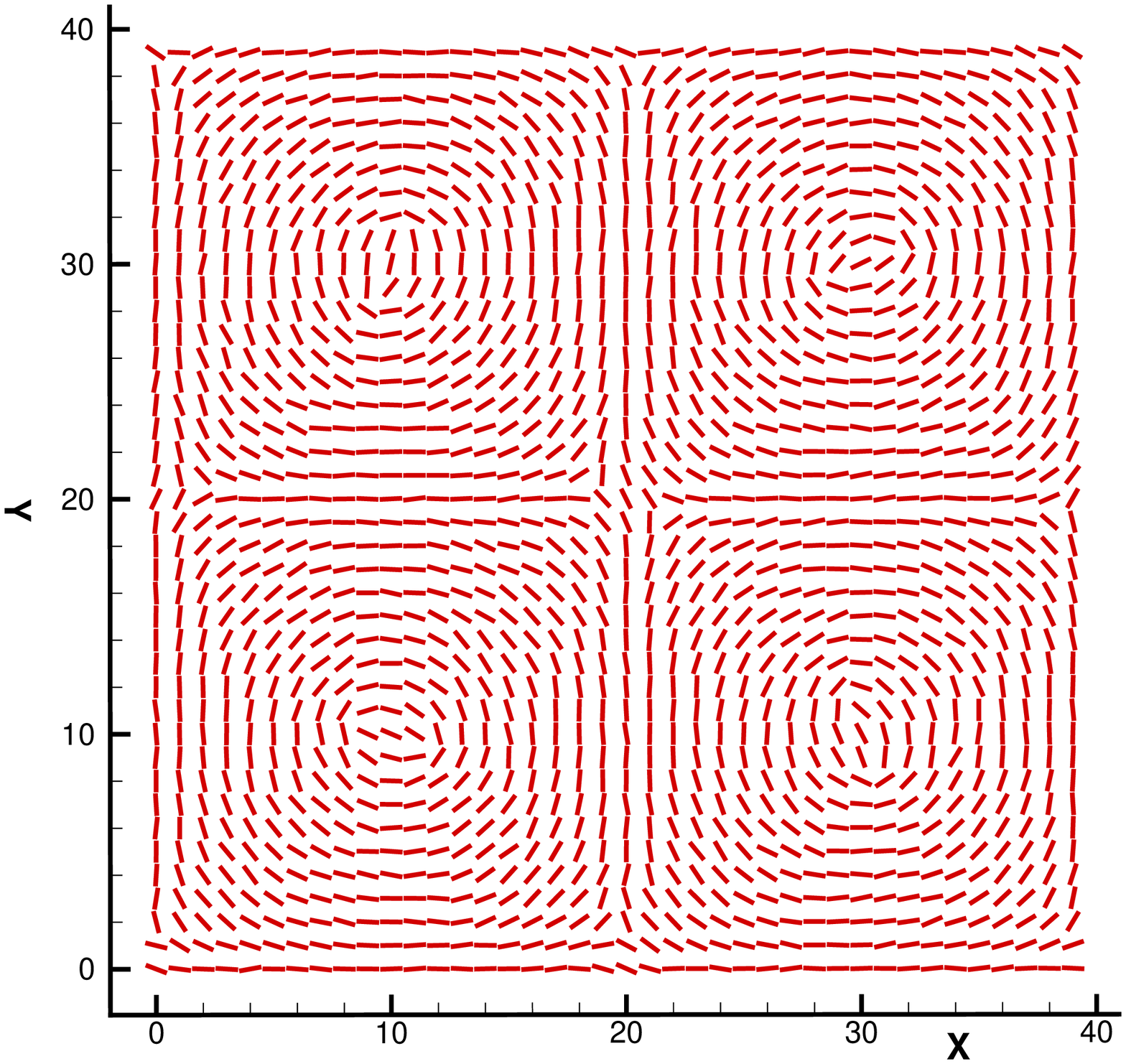}
\caption{(Color online) Director patterns in three layers of a cell of size $L=40$, $H=6$ with defect lattice spacing $a=20$. The two identical patterned layers ($z=0,5$) are displayed in (a). Layers $z=1$ and $z=3$, shown in (b) and (c) respectively, illustrate the bifurcation of the integer--valued point defects on the patterned surfaces into pairs of half--integer defect lines which traverse the cell in the $z$ direction, joining integer defects on opposite sides of the cell (see Fig.~\ref{smallHdefects}). }
\label{smallH}
\end{figure}

\begin{figure}

\includegraphics[height=3in]{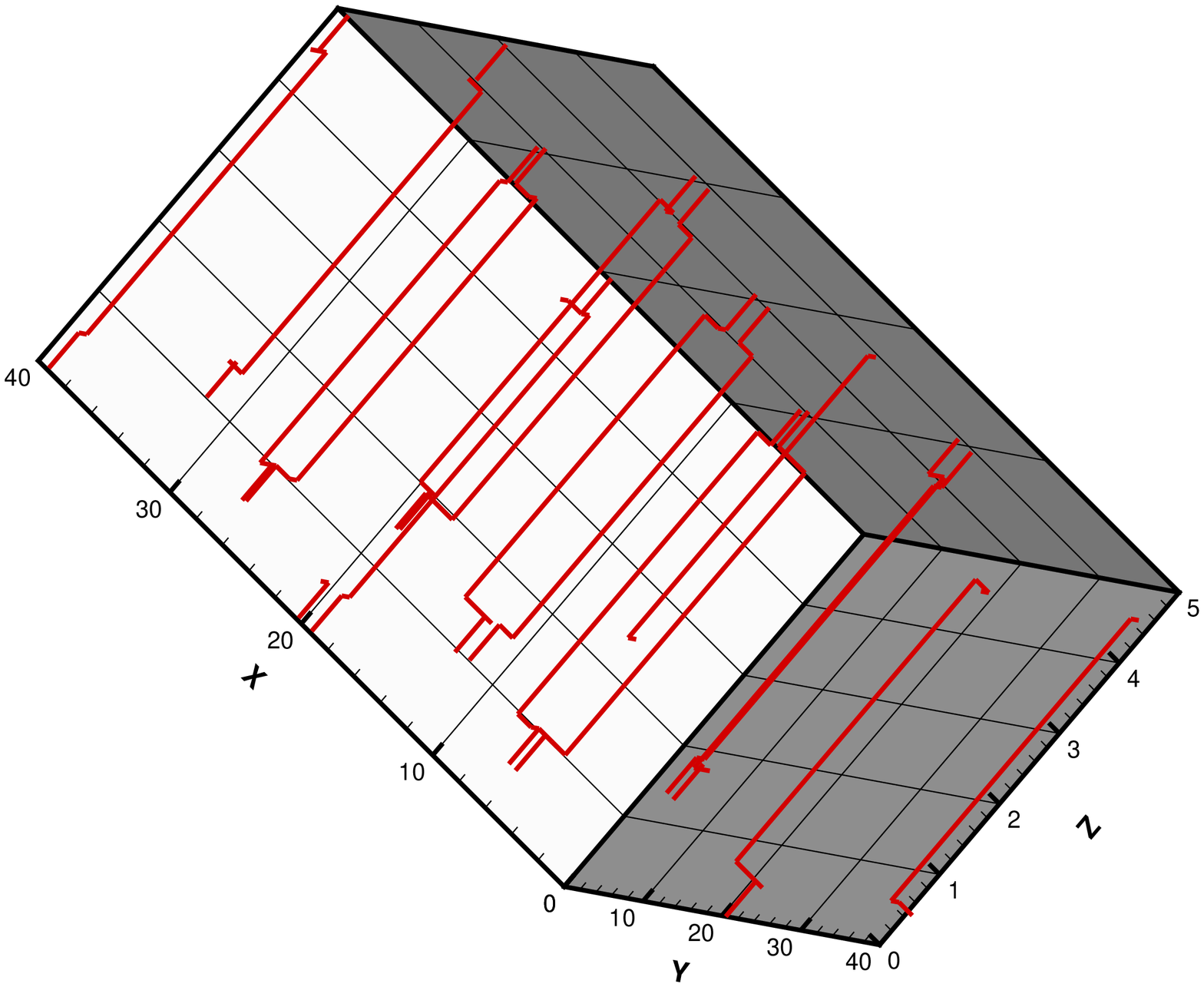}
\caption{(Color online) Full cell view of the half--integer valued defect lines in a cell of size $L=40$, $H=6$ with defect lattice spacing $a=20$. The defect lines emerge from the point defects on the patterned boundary layers located at $z=0,5$ and traverse the cell joining like--charged point defects on opposite sides.}
\label{smallHdefects}
\end{figure}

We carried out simulations for $a=10, 20, 22$ and $H=4,6, 8, 10, 12, 14, 16, 17, 18, 19, 20, 40$, with selected combinations of these parameters yielding values of the ratio $H/a$ ranging from 0.2 to 4.  We chose $L=40$ so that our simulation box contains at least four unit cells of the pattern (though we did do some runs with $L$ as large as 80 to check the robustness of our results). We considered several ways of modeling the core of the defects, either placing a rotor pointing in the $z$ direction at the center of the core, or arranging the rotors of the Lebwohl--Lasher lattice so that the core is centered on a plaquette of the lattice, i.e., there are no lattice sites allocated to the core. Our key results are independent of how we model the core.  Our results indicate that the equilibrium state of the system depends crucially on the ratio $H/a$:
\begin{enumerate}
\item[(i)] When $H/a \gtrsim 0.9$ we find that the system is very well--ordered (with nematic order parameter  $S \simeq 0.95$) along one of the two diagonal directions in the $x-y$ plane (the two diagonal directions are equivalent and the system chooses one or the other depending on the seed of the Monte Carlo algorithm). The nematic order is locally very uniform for $1 < z < H-2$ as illustrated in Fig.~\ref{largeH}. The $\pm 1$ defect points on the patterned surfaces bifurcate into half--integer defect lines (identified using the method of Ref.~\cite{Zapotocky:95}) which hug the patterned surfaces as indicated in Fig~\ref{largeHdefects}.  As shown in the figure a half--integer line emerging from a +1 defect on the surface links up with one of the half--integer lines emerging from a neighboring -1 defect on the \textit {same} patterned surface (note that $+1/2$ and $-1/2$ defect lines are topologically equivalent in a nematic), thus forming diagonal chains of half--integer defect lines. The nematic director undergoes a $90^\circ$ rotation as a half--integer line is crossed. In this way the defect lines mediate the transition of the director orientation from uniform diagonal order within the center of the cell to the patterned array of integer defects on the alignment layers.
\item[(ii)] When $H/a  \lesssim 0.9$ the nematic order parameter vanishes. The defect points on the surfaces once again bifurcate into half--integer lines in the interior of the cell but these lines now thread through the cell, i.e., a half--integer line from a +1 defect on one surface links up with a line emerging from the +1 defect located on the other surface at the same point in the $x-y$ plane (the -1 defects are linked in a similar fashion). See Figs.~\ref{smallH} and \ref{smallHdefects}. We do not see any evidence of integer--valued lines threading through the sample. Only pairs of half--integer lines which have bifurcated from the $\pm 1$ defects on the surfaces are present~\cite{Andrienko-Allen}, and we found no evidence of ``escape in the third dimension'' \cite{Meyer:73,Cladis:72} in these samples; the rotors have negligible tilt out of the $x-y$ plane.
 \end{enumerate}

We bounded our estimate of the critical value of $H/a$ by simulating systems with $H=20, a=22, H/a=0.91$ and $H=18, a=20, H/a=0.90$. The former yields behavior (i) above, while the latter yields behavior (ii). 


The significance of the value of the ratio $H/a$ in determining the equilibrium state can be understood in very simple physical terms. If a defect line crosses the cell parallel to the $z$ axis, joining defects of like charge on opposite patterned surfaces, then the defect line length will be of order $H$ (it will be exactly $H$ if the line is perfectly straight). If the line instead hugs the surface as in Fig.~\ref{largeHdefects} and links up with a line emerging from a neighboring defect point of opposite charge, then the line length will be of order $a$ \cite{footnote1} (exactly $a$ if the line does not wander). This simple argument suggests that the critical value of the ratio $H/a$ should be 1; our simulations yield a value of approximately 0.9.  

\begin{figure}
\begin{center}
\includegraphics[height=3in]{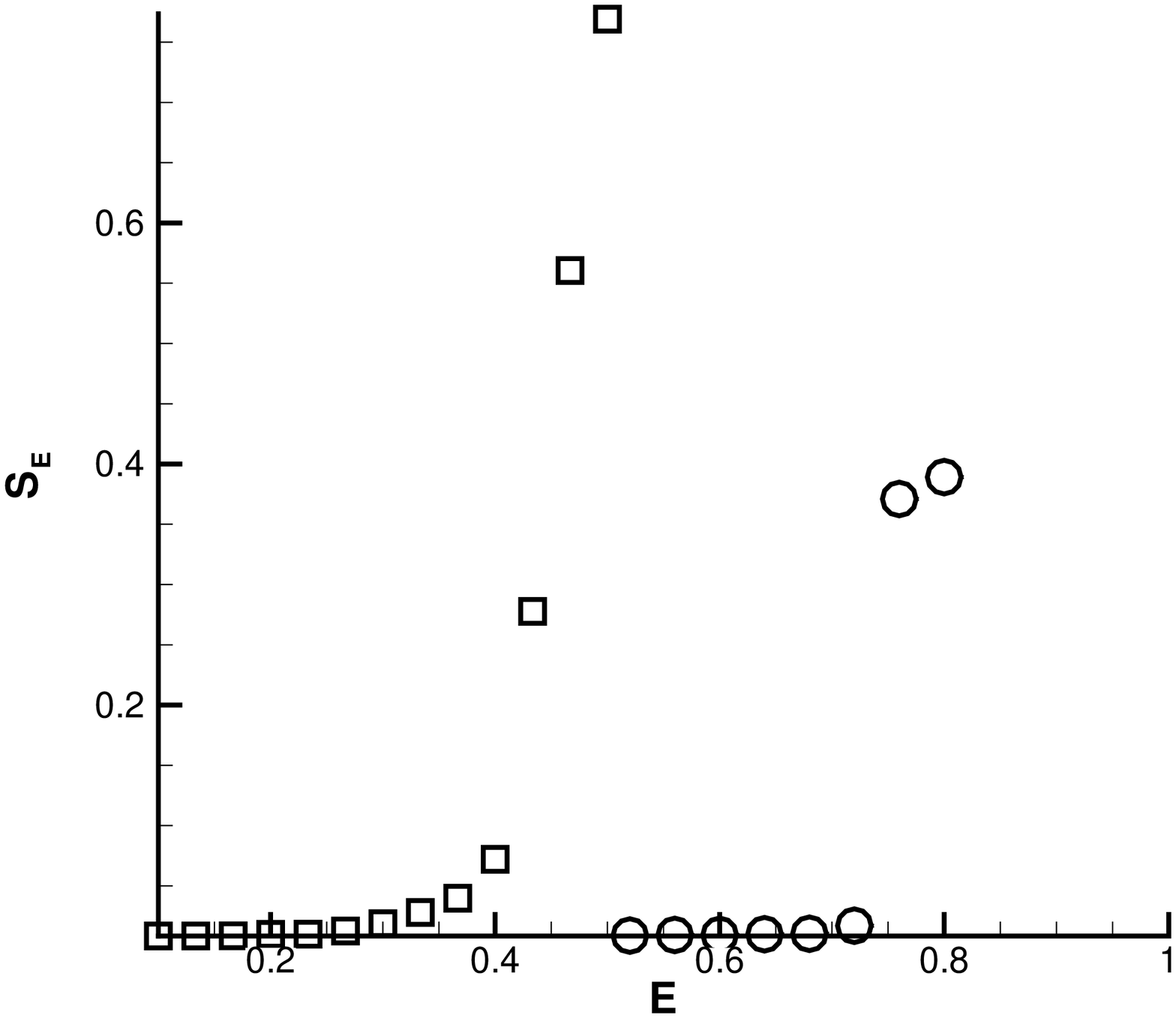}
\caption{}
\label{field}
\end{center}
\end{figure}

We also studied the switching behavior of the cell in the presence of an external electric field pointing in the $z$ direction. In the absence of the field the rotors have negligible tilt out of the $x-y$ plane. A sufficiently strong field will cause the rotors to tilt and eventually order in the $z$-direction within the interior of the cell. To study this field--induced transition we added an additional contribution $\cal {H}_E$ to the Hamiltonian: 

\begin{equation}
{\cal {H}_E}= -\Delta \epsilon \epsilon_0 \sum _{i}
({\mathbf E} \cdot  {\mathbf s}_{i})^2, 
\end{equation}
where $\Delta\epsilon$ is the dielectric anisotropy of the nematic, $\epsilon_0$ is the permittivity of vacuum, and $\mathbf E = E \mathbf {\hat z}$. In Fig.~\ref{field} we display the nematic order parameter $S_E$ measured with respect to the field direction (the $z$ axis):

\begin{equation}
S_E=\frac {1}{N}\sum_i \biggl\lbrace {3 \over 2}
({\mathbf s}_{i} \cdot  {\mathbf {\hat z}})^2 -{1\over
2}\biggr\rbrace
\end{equation}
for two cells, one of height $H=6$, the other with $H=40$ (both have the same lateral dimension $L=40$ and pattern lattice size $a=20$). In both cases there is a nonzero threshold for switching with the value of the critical field decreasing with increasing height of the cell. For a given value of $H$ we found that the value of the critical electric field was insensitive to the value of the lattice constant $a$.

\begin{figure}
\includegraphics[height=3in]{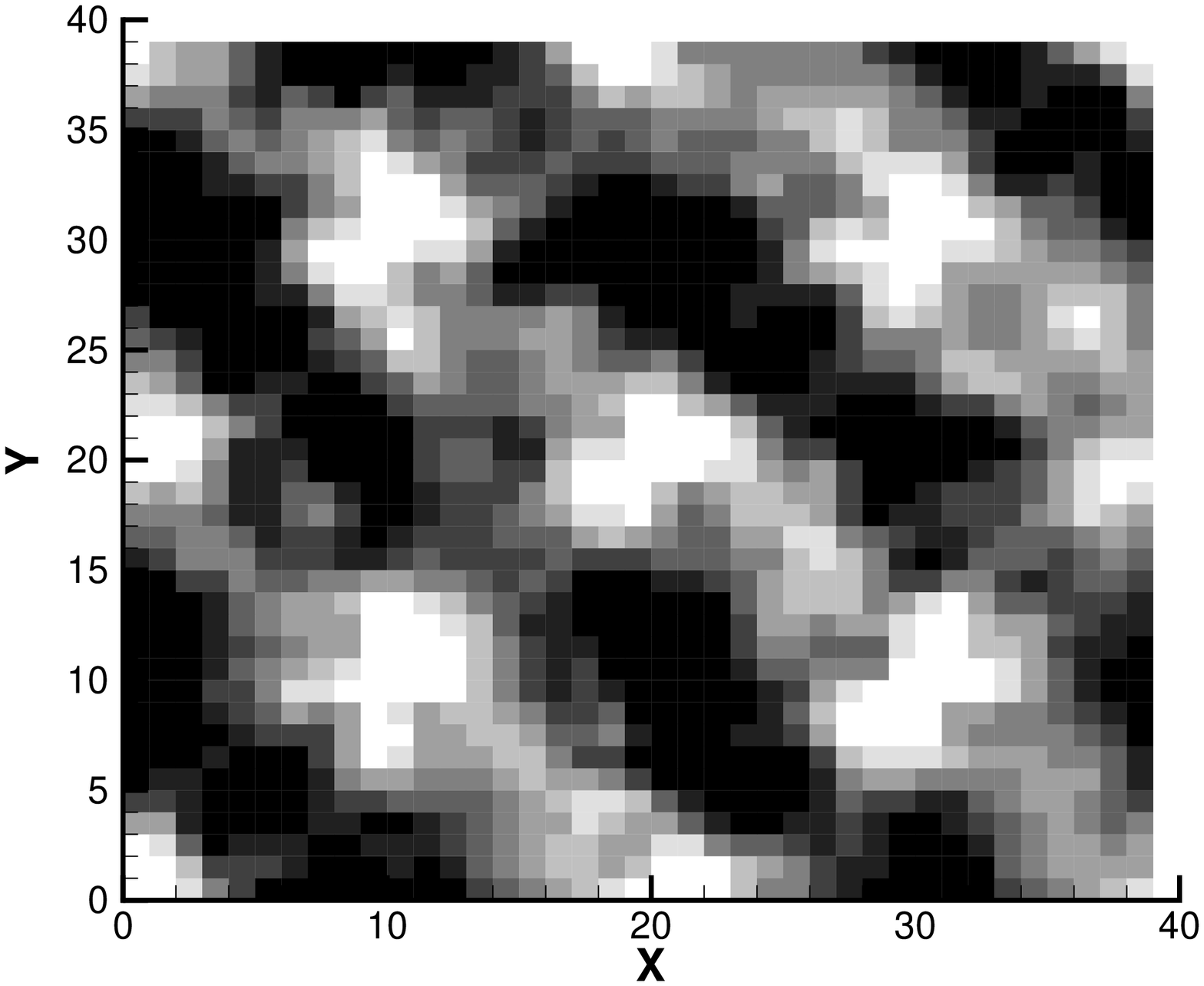}
\includegraphics[height=3in]{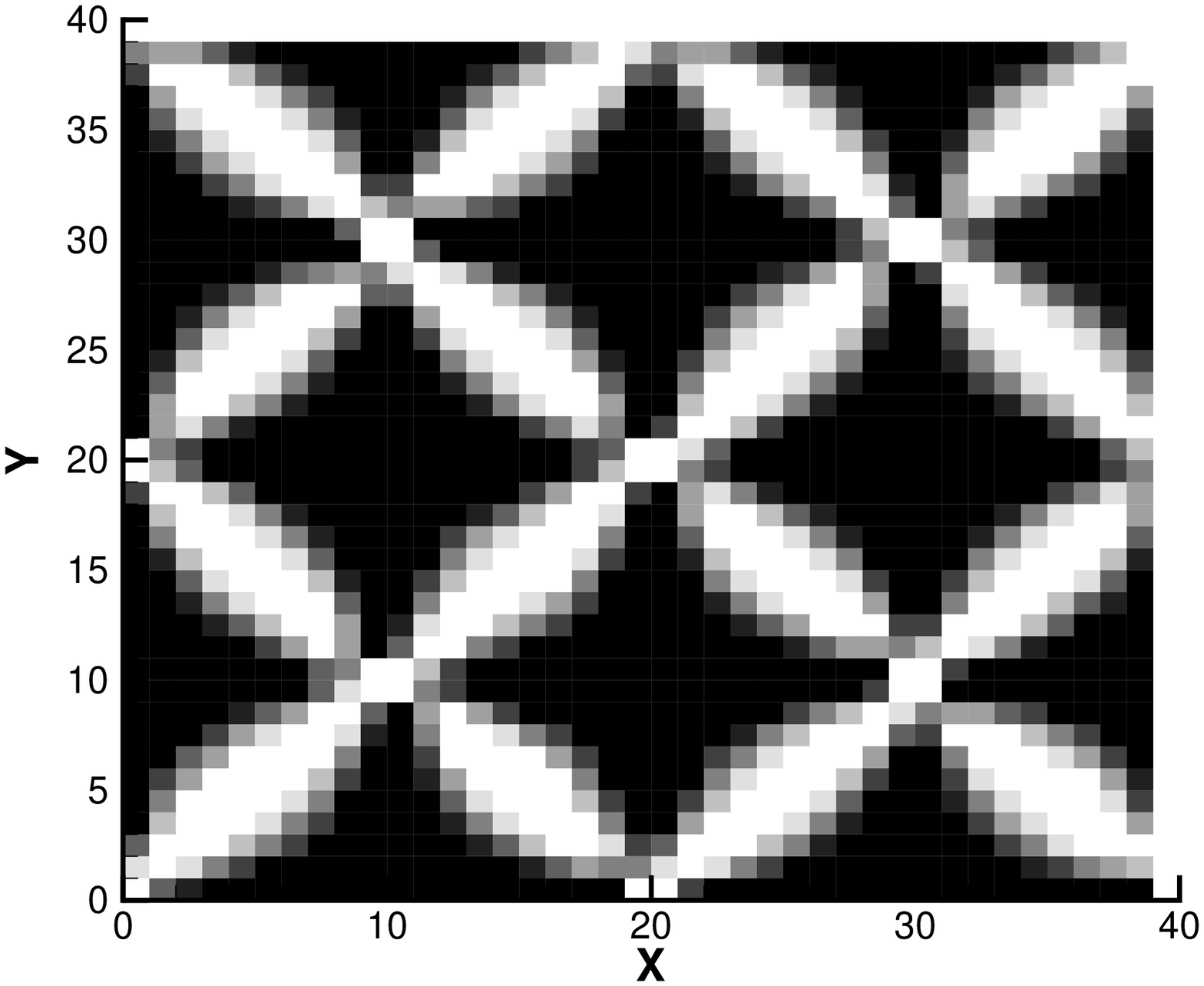}
\caption{Simulated polarized optical textures for two nematic cells: (a) $H=40$ (see Figs.~\ref{largeH} and \ref{largeHdefects}) and (b) $H=6$ (see Fig.~\ref{smallH} and \ref{smallHdefects}), both with $L=40$ and defect lattice spacing $a=20$. }
\label{optical}
\end{figure}

An eventual comparison of our results requires information about the transmission of light through the cell. To this end, we simulated polarized optical textures using the M\"uller matrix technique \cite{Ondriscrawford:91,Berggren:94}. We simulated a pair of crossed polarizers, one on either side of the cell, one along the $x$ axis, the other along the $y$ axis. In Fig.~\ref{optical} we display results for the two systems shown in Figs.~\ref{largeH}--\ref{smallHdefects}.  These results are consistent with the descriptions of the nematic ordering and defect line geometry described above for the two states, one with $H/a>0.9$, the other with $H/a <0.9$. In the former case (Fig.~\ref{optical}a) we see a diagonal light pattern corresponding to the strong nematic ordering along a diagonal direction in the $x-y$ plane, with the bright stripes in the pattern oriented \textit {perpendicular} to the direction of the nematic order. In the latter case, corresponding to Fig.~\ref{optical}b, the defect lines thread through the cell. Because the pair of half--integer lines joining two integer--valued defects are closely spaced together they appear in the optical pattern like a +1 defect with a large core.

\section{Conclusions}
\label{conclusions}
Using Monte Carlo simulations of the Lebwohl--Lasher model we have studied nematic order in a cell whose top and bottom layers are patterned with a regular array of $\pm 1$ point topological defects. We found that the nematic order in the cell depends crucially on the ratio of the height of the cell $H$ to the lattice spacing $a$ of the defects. When $H/a \gtrsim 0.9$ the system is very well--ordered in spite of the frustration induced by the lattice of defects. This frustration is relieved by a network of half--integer defect lines which emerge from the point defects and hug the top and bottom surfaces of the cell. Within the interior of the cell the director is remarkably uniform and points along one of the diagonal directions in the $x-y$ plane. When $H/a \lesssim 0.9$ the system is disordered and the half--integer defect lines thread through the cell joining point defects on opposite surfaces. The dependence of the ordering on the ratio $H/a$ and its approximate value can be understood in terms of minimization of the length of the defect lines joining the point defects. An individual defect line has length approximately $H$ if the line traverses the cell from top to bottom and approximately $a$ if it joins defects on the same surface (the length will be exactly these values if the line is as straight as possible). Thus, when $H\gtrsim a$ the system can minimize the defect line energy by joining defects on the same surface, whereas if $H \lesssim a$, the defect line energy is minimized by having the lines traverse the height of the cell. 

The Lebwohl--Lasher model assumes a single Frank elastic constant proportional to the coupling $J$. When the defect lines traverse the height of the cell for small $H$, the defect energy is determined primarily by the bend elastic constant, $K_3$, whereas when the lines hug the surfaces for large values of $H$, it is the twist elastic constant $K_2$ which determines the defect energy. Thus, in a more realistic model where $K_2<K_3$ \cite{deGennesbook}, we might expect that the critical value of $H/a$ will be smaller than the value of 0.9 found here for the Lebwohl--Lasher model.  

It is experimentally feasible \cite{Crawford:private} to fabricate a cell with alignment layers of the form considered here. While it is not easy experimentally to vary $H$, the cell gap, it is possible to vary $a$ by suitable manipulation of the optical beams used to create the interference pattern. Preliminary estimates suggest that it should be possible to vary $a$ from 1 to 3 $\mu m$. With a typical cell gap $H$ of 2 $\mu m$ this would allow $H/a$ to range from 2 to 0.66. Assuming that our estimate 0.9 of the critical value of this ratio from the Lebwohl--Lasher model is reasonably close to the value for real materials, we would expect that the behavior shown in Fig.~\ref{optical} could be seen experimentally.

\section*{Acknowledgments}
We are grateful to G. Crawford, J. Eakin and S. Gorkhali for very helpful discussions which motivated our work.

%\bibliography{refs.bib}

\begin{thebibliography}{24}
\bibitem[{foo()}]{footnote1}
\bibinfo{note}{The length is $a$ rather than $a/\sqrt{2}$ because the defect
  lines in a lattice model are made up of straight line segments parallel to
  either the $x$ or $y$ axis; see Fig.~\ref{largeHdefects}.}
\end{thebibliography}

\end{document}